\documentclass[prd,twocolumn,amsmath]{revtex4}
\usepackage{amssymb,amsmath}
\usepackage{graphicx}
\usepackage{color}
\usepackage{graphicx}

\usepackage[dvipsnames]{xcolor}

\begin{document}
\title{Scattering Cross Section and Stability of Global Monopoles}

\author{J. P. M. Pitelli}
\email[]{pitelli@ime.unicamp.br}
\affiliation{Departamento de Matem\'atica Aplicada, Universidade Estadual de Campinas, 13083-859, Campinas, S\~ao Paulo, Brazil}
\author{V. S. Barroso}
\email[]{barrosov@ifi.unicamp.br}
\affiliation{Instituto de F\'isica ``Gleb Wataghin'', Universidade Estadual de Campinas, 13083-859, Campinas, S\~ao Paulo, Brazil}
\author{Maur\'icio Richartz}
\email[]{mauricio.richartz@ufabc.edu.br}
\affiliation{Centro de Matem\'atica, Computa\c{c}\~ao e Cogni\c{c}\~ao, Universidade Federal do ABC (UFABC), 09210-170 Santo Andr\'e, S\~ao Paulo, Brazil}

\begin{abstract}

We study the scattering of scalar waves propagating on the global monopole background. Since the scalar wave operator in this topological defect is not essentially self-adjoint, its solutions are not uniquely determined until a boundary condition at the origin is specified. As we show, this boundary condition manifests itself in the differential cross section and can be inferred by measuring the amplitude of the backscattered wave. 
We further demonstrate that whether or not the spacetime is stable under scalar perturbations also relies on the chosen boundary condition. In particular, we identify a class of such boundary conditions which significantly affects the differential cross section without introducing an instability.

\end{abstract}

\maketitle

\section {Introduction}

Topological defects are formed during phase transitions in the early universe. They originate from the breakdown of gauge symmetries and are believed to seed the formation of large-scale structure in the universe~\cite{vilenkin1,vilenkin2}. Depending on which symmetry is broken, they are classified as domain walls, cosmic strings or global monopoles~\cite{vilenkin2}. Global monopoles, in particular, arise when the global $O(3)$ symmetry of the Lagrangian 
\begin{equation} \label{lagrangian}
\mathcal{L}=\frac{1}{2}\partial_\mu\phi^{a}\partial^{\mu}\phi^{a}-\frac{1}{4}\lambda\left(\phi^{a}\phi^{a}-\eta^2\right)^2,
\end{equation}
where $\phi^a$ ($a=1,2,3$) is a triplet of scalar fields,
is spontaneously broken to $U(1)$~\cite{barriola}.

The metric around a global monopole, once its core size has been neglected, can be approximated by
\begin{equation}
ds^2=-dt^2+dr^2+\alpha^2r^2 \left(d\theta ^2 + \sin^2 \theta d \phi^2 \right),
\label{metric monopole}
\end{equation}
where  the parameter $\alpha=1-8\pi G\eta^2$ depends on the symmetry breaking energy scale (typically $8\pi G\eta^2\approx 10^{-5}$ in grand unified theories). This metric describes a spacetime with a deficit solid angle (the section $\theta=\pi/2$ coresponds to  a cone with deficit angle $\Delta= 8\pi^2G\eta^2$). The spacetime is not flat, being characterized by the  curvature scalar $R=2\left(\alpha^{-2}-1\right)r^{-2}$~\cite{bezerra de mello}. The energy density, determined by the $00$-th component of the stress-energy-momentum tensor $T_{\mu \nu}$, is given by $T_{00}\sim G\eta^2/r^2$ so that the total energy $E(r)\sim 4\pi G \eta^2 r$ is linearly divergent for large $r$. Despite the fact that the Ricci scalar goes to zero when $r\to\infty$, the global monopole is not asymptotically flat since there are non-zero components of the Riemann curvature tensor $R_{\rho \sigma \mu \nu}$ for arbitrarily large $r$. In particular, the $R_{\theta \phi \theta \phi}=\left(1-\alpha^2\right)\sin^2{\theta}$ component is non-zero if $\alpha \neq 1$. In this paper, we will consider scattering on the global monopole spacetime. We argue that our results are valid in the $\alpha\lessapprox 1$ (i.e.~$\Delta \ll 1$) regime, which is the realistic one predicted by grand unified theories.

The propagation of a massless scalar field $\Psi$ around the global monopole background is governed by the Klein Gordon equation, $(\nabla _ \nu \nabla ^\nu - \mu^2 )  \Psi = 0$. Its solutions, however, are not uniquely determined by the initial data. In fact, if the spatial part of the wave equation is seen as an operator $A$ acting on a certain $L^2$ Hilbert space, an infinite number of sensible dynamical prescriptions may be defined, each one corresponding to a different choice of a self-adjoint extension for $A$~\cite{wald}. These various extensions are encoded in the arbitrary specification of a boundary condition at $r=0$.

According to Horowitz and Marolf~\cite{horowitz}, a classically singular spacetime is said to be quantum mechanically singular when the evolution of a wave packet on the spacetime background depends on extra information not predicted by the theory. In this sense, the evolution of a wave packet in the global monopole spacetime is as uncertain as the evolution of a classical particle due to the presence of the classical singularity at $r=0$. Even though the chosen boundary condition cannot be directly observed, we should expect that some physical observable quantities will depend on it. The phase difference between incident and scattered waves is an example of that~\cite{mosna}, but in this paper we will focus on the differential  scattering cross section. As we show, this cross section is not determined until we specify a boundary condition. Stated in another way, one could use observable information obtained from a scattering experiment (i.e.~the cross section) to determine the boundary condition favored by Nature. 

In Ref.~\cite{mazur}, the scattering of scalar waves by a global monopole was analyzed for a Dirichlet boundary condition and an approximation for the total cross section was obtained. More recently, in Ref.~\cite{anacleto}, Anacleto et al.~considered the scattering of scalar waves by a black hole with a global monopole and showed that the differential cross section for small angles contains explicitly the $\alpha$ parameter of the global monopole. Here, on the other hand, we consider not only a Dirichlet boundary condition (which is usually assumed since it leads to regular solutions at the origin), but all possible boundary conditions allowed by self-adjointness. We investigate how much the differential cross section of the global monopole for scalar waves depends on the choice of the boundary condition. 

To be physically relevant, a spacetime should be stable (or, if unstable, should have an instability time scale small enough compared to the time scales of the  effects under investigation). Because of that, we also study the stability of the global monopole. Similar work was done in Refs.~\cite{perivolaropoulos,achucarro}. In Ref.~\cite{perivolaropoulos}, for instance, it was demonstrated that the global monopole is stable under a radial rescaling $r\to\kappa r$. That is, if we impose a cutoff $r=R_c$ in a cosmological setup, the energy $E(R_c)$ has a minimum at $\kappa=1$. In Ref~\cite{achucarro}, it was shown that the global monopole is stable under axisymmetric perturbations of the triplet $\phi^a$. In this paper, on the other hand, we follow a different approach by considering perturbations of a scalar test field. Encoding the arbitrary boundary condition as a free parameter, we search for solutions of the scalar wave equation which correspond to unstable modes, i.e.~purely outgoing modes at spatial infinity that grow exponentially in time.
   
Our work is organized as follows:  in Sec.~\ref{secii}, we briefly review the necessity of choosing a boundary condition to solve the wave equation in the global monopole spacetime. We follow Ref.~\cite{pitelli}, where the singular nature of the global monopole spacetime was analyzed and the authors found that a Robin boundary condition is necessary to make the spatial part of the wave operator self-adjoint. Next, in Sec.~\ref{Sec. Scattering}, we use the method of partial waves to find the differential cross section for scalar waves and investigate its relation to the boundary condition. In Sec.~\ref{seciv} we analyze the stability of the global monopole under scalar perturbations of a test field and show that the spacetime is stable for a class of boundary conditions. The last part, Sec.~\ref{concluions}, is reserved for our final considerations.

\section{Boundary Conditions for the Klein-Gordon Equation}
\label{secii}
Consider a massless scalar field $\Psi$ propagating on the global monopole background. (The massive case, discussed in Ref.~\cite{pitelli}, only brings unnecessary complications). The associated Klein-Gordon equation, in view of \eqref{metric monopole}, can be cast as
\begin{equation}\begin{aligned}
\frac{\partial^2\Psi}{\partial t^2}=&\frac{1}{r^2}\left(r^2\frac{\partial^2\Psi}{\partial r^2}\right)+\frac{1}{\alpha^2\sin{\theta}}\frac{\partial}{\partial \theta}\left(\sin{\theta}\frac{\partial \Psi}{\partial \theta}\right)\\&+\frac{1}{\alpha^2r^2\sin^2{\theta}}\frac{\partial^2\Psi}{\partial \varphi^2}.
\label{complete equation}
\end{aligned}\end{equation}
Due to the spherical symmetry, the above equation  is separable under the \textit{ansatz} $\Psi(t,r,\theta,\varphi)=R_{\omega\ell}(r)Y_{\ell}^{m}(\theta,\varphi)e^{-i \omega t}$, where $Y_\ell^m(\theta, \varphi)$ are the usual spherical harmonics, $\ell \in \mathbb{N}$ is the orbital quantum number, $m \in \mathbb{Z}$ ($-\ell \le m \le \ell$) is the azimuthal number, and $\omega \in \mathbb{C}$ is the complex wave frequency. A straightforward calculation transforms Eq.~\eqref{complete equation} into an equation for the radial function $R_{\omega\ell}$,
\begin{equation}
R_{\omega \ell}''(r)+ \frac{2}{r} R_{\omega \ell}'(r)+\left[\omega^2-\frac{\ell(\ell+1)}{\alpha^2 r^2}\right]R_{\omega \ell}(r)=0.
\label{r-equation}
\end{equation}

Note that scalar waves are only affected by the $\alpha$ parameter of the global monopole through the inverse square potential $V_{\ell}(r)=\ell(\ell+1)/(\alpha^2r^2)$. In other words, only non-zero ($\ell \neq 0$) angular momentum waves will perceive the angular deficit. Spherical waves ($\ell = 0$), on the other hand, are unaffected by the parameter and will propagate as in Minkowski spacetime. Therefore, the true classical singularity at $r=0$ will be perceived only by $\ell = 0$ waves since, for $\ell \neq 0$ waves, it becomes ``invisible" due to the strong repulsive potential.  

Let us now understand how the remark above translates into the necessity of a boundary condition for spherical waves. The general solution of Eq.~\eqref{r-equation} is simply
\begin{equation} \label{eqn1}
R_{\omega \ell}(r)=A_{\omega \ell}\frac{J_{\nu_\ell}(\omega r)}{\sqrt{\omega r}}+B_{\omega \ell}\frac{N_{\nu_\ell}(\omega r)}{\sqrt{\omega r}},
\end{equation}
where $A_{\omega \ell}$, $B_{\omega \ell}$, and $\nu_\ell=\frac{1}{2}\sqrt{1+\frac{4\ell(\ell+1)}{\alpha^2}}$ are constants, and $J_\nu(\omega r)$ and $N_{\nu}(\omega r)$ are the $\nu$-th order Bessel and Neumann functions, respectively. Note that if we restrict the frequency to be real, i.e.~$\omega \in \mathbb{R}$, the function $J_{\nu_\ell}(\omega r)/\sqrt{\omega r}$ is square-integrable near the origin for all $\ell \in \mathbb{N}$:
\begin{equation}
\int_{0}^{c}{\left|\frac{J_{\nu_\ell}(\omega r)}{\sqrt{\omega r}}\right|^2r^2dr}<\infty,
\end{equation}
where $c$ is an arbitrary positive constant.
On the other hand, the function $N_{\nu_\ell}(\omega r)/\sqrt{\omega r}$, with $\omega \in \mathbb{R}$, is square-integrable near the origin only for $\ell=0$. In view of that, to avoid non-square-integrable solutions, the boundary condition $B_{\omega \ell}=0$ naturally arises for $\ell \neq 0$ waves. For $\ell=0$, since both solutions are square-integrable, an arbitrary boundary condition at $r=0$ must be chosen. 

It is important to remark here that, even though the wave equation is the same for the Minkowski spacetime and for the global monopole (when $\ell=0$), there is a crucial difference between the two cases. In the first one, the origin is not a singularity of the spacetime. Consequently, the coefficient $B_{\omega0}$ must also vanish since the Laplacian of $N_{\nu_0}(\omega r)/\sqrt{\omega r}$ is proportional to the Dirac delta function $\delta^3(r,\theta,\phi)$, which fails to be square-integrable~\cite{ishibashi}. The global monopole, however, has a singularity at the origin $r=0$, which is not part of the manifold. As a result, $N_{\nu_0}(\omega r)/\sqrt{\omega r}$ is square-integrable and the mode $\ell = 0$ is allowed.

It is convenient to define a new radial function $G_{\omega \ell}(r)=r R_{\omega \ell}(r)$ so that, in terms of $G_{\omega \ell}$, Eq.~\eqref{r-equation} becomes
\begin{equation}
\frac{d^2G_{\omega \ell}(r)}{dr^2}+\left[\omega^2-V_{\ell}(r)\right]G_{\omega \ell}(r)=0.
\label{equation simplified0}
\end{equation}
Another way to understand why a boundary condition is needed when $\ell\neq 0$ is that the inverse square potential $V_{\ell}(r)$ falls off faster than $3/4r^2$, which is a well-known requirement for having a function which is not square-integrable~\cite{gitman}. When $\ell=0$, the repulsive potential is absent, and the equation above reads
\begin{equation}
\frac{d^2G_{\omega 0}(r)}{dr^2}+\omega^2G_{\omega 0}(r)=0.
\label{equation simplified}
\end{equation}

The most general boundary condition for $G(r)$ is the Robin mixed boundary conditions (see Refs.~\cite{ishibashi,mosna}),
\begin{equation}
G_{\omega 0}(0)+\beta G_{\omega 0}'(0)=0,
\label{boundary condition}
\end{equation}
where $\beta\in\mathbb{R}\cup \{-\infty,+\infty\}$ is an arbitrary parameter. When this boundary condition is taken into account, the solution of \eqref{equation simplified}, written in terms of the parameter $\beta$, becomes
\begin{equation}
G_{\omega 0}^\beta(r)\sim \left\{\begin{aligned}&\sin{(\omega r)}-\beta \omega\cos{(\omega r)},&\textrm{for}~\beta\neq \pm \infty,\\
&\cos{(\omega r)},&\textrm{for}~\beta=\pm\infty.
\end{aligned}\right.
\label{Gbeta}
\end{equation}

To the best of our knowledge, all previous work on scattering by the global monopole spacetime assumed a Dirichlet boundary condition ($\beta = 0$), which does not allow for the existence of bound states. For some other values of the boundary condition parameter $\beta$, however, bound states do exist. In fact, if we let the frequency $\omega$ be imaginary so that $\omega^2=-\lambda^2<0$, the general solution of Eq.~\eqref{equation simplified} becomes   
\begin{equation}
G_{\lambda 0}(r) = C_{\lambda 0}e^{-\lambda r} + D_{\lambda 0} e^{\lambda r},
\end{equation}
where $C_{\lambda 0}$ and $D_{\lambda 0}$ are constants (without loss of generality we can assume $\lambda>0$). Since we are looking for square-integrable solutions, we must have $D_{\lambda 0}=0$. In such a case, the boundary condition (\ref{boundary condition}) transforms into \begin{equation}
C_{\lambda 0}(1-\lambda\beta)=0.
\label{equation for lambda}
\end{equation}
In order to have non-trivial solutions, the parameters must be related by $\lambda=1/\beta$, which only makes sense when $\beta>0$ (otherwise $\lambda$ would be negative). The associated solution is then the bound state 
\begin{equation}
R_{\textrm{bound}}(r)\sim \frac{e^{-r/\beta}}{r}.
\label{Rbound}
\end{equation}
When $\beta = 0$ or $\beta=\pm\infty$, it is straightforward to show that no bound states are allowed.

\section{Wave Scattering}
\label{Sec. Scattering}

In this section we study the scattering of incident scalar waves  on the global monopole. Using the method of partial waves, our main goal is to determine the differential cross section of the global monopole as a function of the boundary condition parameter $\beta$. We reemphasize that our results must be applied to the realistic case considered in grand unified theories $8\pi G\eta^2\approx 10^{-5}$ ($\alpha\lessapprox 1$), where $R_{\theta\phi\theta\phi}\approx 10^{-5}\sin^2{\theta}$ . However, even for such small angular defects, the equatorial plane  corresponding to a cone with a very small deficit angle. Scattering in conical spacetimes was discussed in Refs.~\cite{deser,barroso}. In~\cite{deser} the authors showed that even though the partial wave analysis in conical spacetimes leads to divergences, it is possible to redefine the incident wave in order to absorb the divergent terms. In~\cite{barroso}, the authors showed that the same procedure can be done when an arbitrary boundary condition is chosen at the origin. Since the global monopole is plagued with the same problem (the spacetime is not asymptotically Minkowski), we also expect divergences in the partial wave analysis of our work. As we will see, these divergences can be handled by smoothing the singularities of the scattering amplitude [see Eq.~(\ref{scatt_amp_2}) bellow.  Despite the fact that the spacetime is topologically nontrivial, we are able to analyze, at least qualitatively, the scattering of waves satisfying different boundary conditions.

To accomplish that, we consider an incident wave $\Psi_{in} = e^{ikz}e^{-i\omega t}$, with wave number $k=\omega$, propagating along the $z$-axis. It is convenient to expand it into spherical waves using the standard plane wave decomposition,
\begin{equation} \label{solution scattering2}
e^{ikz}=\sum_{\ell=0}^{\infty}{(2\ell+1)i^\ell j_{\ell}(\omega r)P_{\ell}(\cos{\theta})},
\end{equation}
where $j_{\ell}(\omega r)$ is the $\ell$-th order spherical Bessel function and $P_{\ell}(\cos \theta)$  is the $\ell$-th order Legendre polynomial.

 This incident plane wave is scattered by the global monopole, so that the total wave can be written as  
\begin{equation}
\Psi=\Psi_{in}+\Psi_{sc},
\label{asymp}
\end{equation}
where $\Psi_{sc}$ corresponds to the scattered part. Far away from the singularity (as $r\to\infty$), this scattered part is an outgoing wave of the form 
\begin{equation} \label{psisc}
\Psi_{sc}=\frac{f(\theta)}{r}e^{ikr}.
\end{equation} 
Similarly, the large-$r$ behaviour of the incident part can be easily determined from Eq.~\eqref{solution scattering2} with the help of the asymptotic expression for the spherical bessel function.  

To determine the scattering amplitude $f(\theta)$, we need the asymptotic behaviour of the solutions we found in the previous section. From Eqs.~\eqref{eqn1} and \eqref{Gbeta}, we find that the general radial solution of the wave equation for an arbitrary parameter $\beta$ is given by
\begin{equation}
R_{\omega \ell}(r)\sim\left\{\begin{aligned}
&\frac{J_{1/2}(\omega r)}{\sqrt{\omega r}}+\beta\omega\frac{N_{1/2}(\omega r)}{\sqrt{\omega r}},&\textrm{for}\,\ell=0,\\
&\frac{J_{\nu_\ell}(\omega r)}{\sqrt{\omega r}},&\textrm{for}\,\ell\neq0.
\end{aligned}
\right.
\end{equation}
Therefore, the full solution \eqref{asymp}, when decomposed into the mode solutions, becomes
\begin{align}
&\Psi(t,r,\theta,\phi)=a_{00}\left(\frac{J_{1/2}(\omega r)}{\sqrt{\omega r}}+\beta\omega\frac{N_{1/2}(\omega r)}{\sqrt{\omega r}}\right)e^{-i \omega t} \nonumber \\
&+ \sum_{\ell = 1}^{\infty} \sum_{m=-\ell} ^{\ell} {a_{\ell m}\frac{J_{\nu_\ell}(\omega r)}{\sqrt{\omega r}}P_\ell(\cos{\theta})e^{i m \phi}e^{-i \omega t}},
\label{complete solution scattering}
\end{align}
where $a_{\ell m}$ are constants to be determined. 

Due to the spherical symmetry, all $m \neq 0$ modes are irrelevant for the scattering, so that $a_{\ell m} = 0$ for them. By comparing the asymptotic behavior of solution~\eqref{complete solution scattering} with the asymptotic behaviour of (\ref{asymp}), we can determine the coefficients $a_{\ell 0}$ to be
\begin{equation}
a_{\ell 0}=\left\{\begin{aligned}
& \sqrt{\frac{\pi}{2}}\frac{i}{i-\beta\omega}, &\textrm{for}\,\ell=0,\\
& \sqrt{\frac{\pi}{2}}(2\ell+1)i^\ell e^{i \delta_{\ell}}, &\textrm{for}\,\ell\neq0,
\end{aligned}\right.
\label{al0}
\end{equation}
where the phase shifts are given by $\delta_\ell = \frac{\pi}{2}\left(\ell+\frac{1}{2}-\nu_\ell\right)$. 
Similarly, the scattering amplitude $f(\theta)$ can be written as,
\begin{equation} \label{scatt_amp1}
f(\theta)= \sum_{\ell=0}^{\infty} \frac{b_{\ell}}{2 i \omega} P_{\ell}(\cos{\theta}),
\end{equation}
where
\begin{equation}
b_{\ell} = \left\{\begin{aligned}
& \frac{2\beta\omega}{i-\beta\omega},   &\textrm{for}\,\ell=0,\\
& (2\ell+1)(e^{2i\delta_\ell}-1), &\textrm{for}\,\ell\neq0.
\end{aligned}\right.
\label{bl0}
\end{equation}

Now, we would like to use the expression above for the scattering amplitude to calculate the differential cross section $d\sigma / d\Omega = |f(\theta)|^2$. However, as in Refs.~\cite{mazur,anacleto}, the infinite sum in \eqref{scatt_amp1} is, depending on the angle $\theta$, either poorly convergent or divergent~\cite{anacleto,cotaescu,yennie}. While nothing can be done for divergent sums, slow convergence can be dealt with with the method described below~\cite{yennie}.   

The first step is to multiply the scattering amplitude by $(1-\cos{\theta})^{n}$, where $n\in\mathbb{N}$, and expand the obtained function in terms of the Legendre polynomials, 
\begin{equation} \label{scatt_amp_2}
(1-\cos{\theta})^n f(\theta)=\frac{1}{2 i \omega}\sum_{\ell=0}^{\infty}{b_\ell ^{(n)}P_\ell(\cos{\theta})},
\end{equation}
where $b_\ell ^{(n)}$ are constant coefficients. By resorting to Bonnet's recursion formula for the Legendre polynomials, it is possible to show that the new coefficients are related to the old ones through $b_\ell^{(n)} = b_\ell$, if $n = 0$, and through the recursive relation
\begin{equation}
b_\ell^{(n)}=b_\ell^{(n-1)}-\frac{\ell+1}{2\ell +3}b_{\ell+1}^{(n-1)}-\frac{\ell}{2\ell-1}b_{\ell-1}^{(n-1)}, 
\label{recursion}
\end{equation}
if $n \ge 1$.  In the end, the scattering amplitude can be written as a so-called reduced series, 
\begin{equation} \label{scatt_amp_3}
f(\theta)=\frac{1}{2i \omega}\sum_{\ell=0}^{\infty}b_\ell^{(n)}\frac{P_\ell(\cos{\theta})}{(1-\cos{\theta})^{n}},
\end{equation}
which converges faster than the series appearing in Eq.~\eqref{scatt_amp1}.

\begin{figure}[h!]\flushleft
\includegraphics[width=0.499\textwidth]{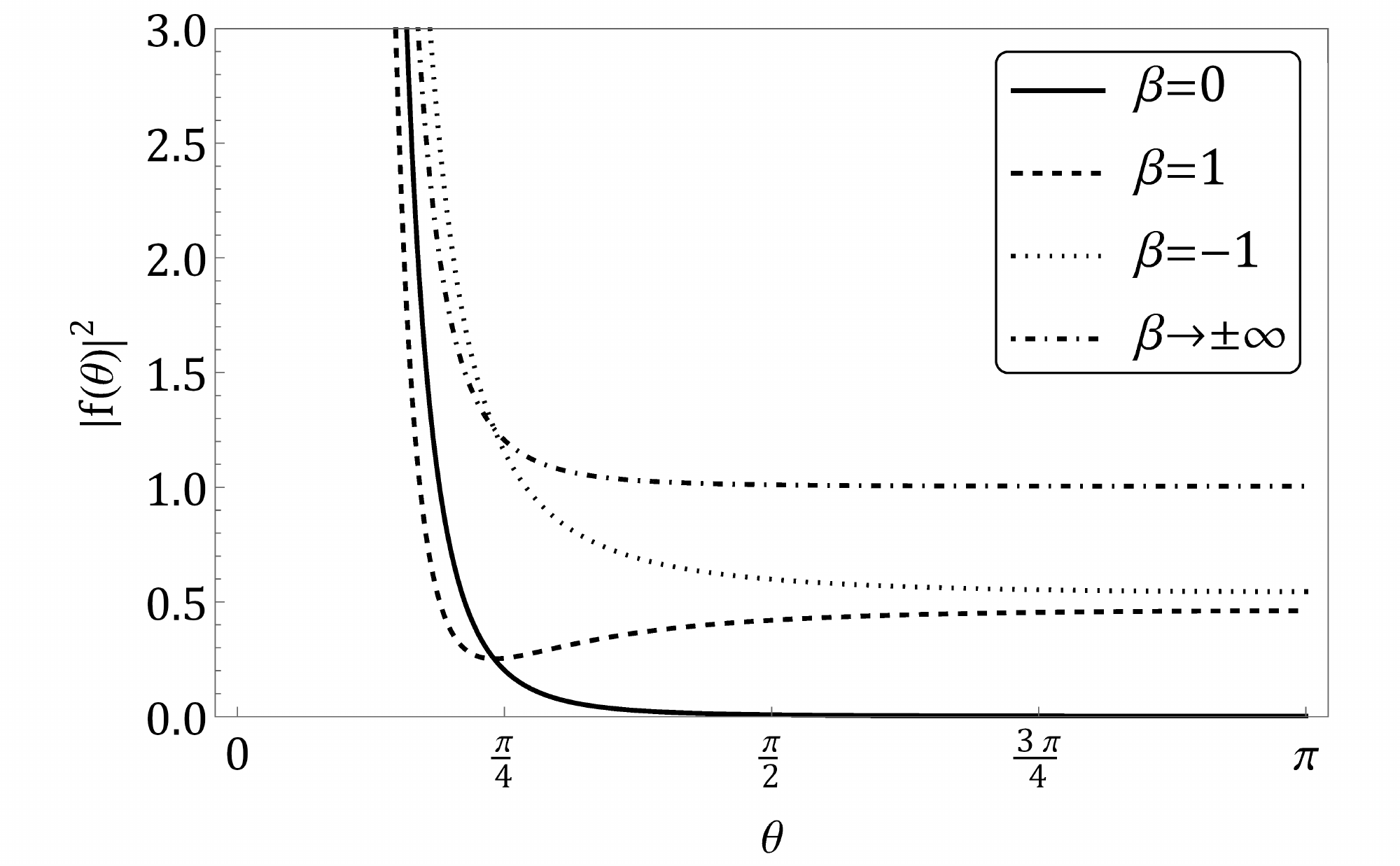}
\caption{Plot of the reduced series of the differential cross section for $\omega=1$, $\alpha=0.95$, and several values of $\beta$.}
\label{graphic differential cross section}
\end{figure}

The last step consists in the numerical implementation of the of the recursive relation~\eqref{recursion}, followed by the calculation of the differential cross section. Using \textit{Mathematica}, we were able to show that taking $n = 6$ is enough to guarantee enough precision when computing the partial sum of the first few terms of the reduced series for $\pi/4 \le \theta \le \pi$, $\omega=\beta=1$. This precision does not seem to change much when different values of $\beta$ and $\omega$ are used. To understand the effect of the boundary conditions on the scattering, we choose $\omega = 1$ and plot in Fig.~\ref{graphic differential cross section}, the differential cross section as a function of the scattering angle $\theta$ for different boundary condition parameters $\beta$. Even though we use an exaggerated $\alpha$ parameter ($\alpha=0.95$) for better  visualization of the effects of the boundary condition, the qualitative behavior is the same for the more realistic value $\alpha\approx 1-10^{-5}$.

The most evident effect of the boundary condition appears to be on the backscattered part ($\pi/2 \le \theta \le \pi$) of the wave. The dirichlet boundary condition, which is usually considered in the literature, produces no backscattering, while the Neumann boundary condition produces the most. We have tested  different values of the frequency $\omega$ and different values of the parameter $\alpha$, and this behavior seems to be universal. As seen in Fig.~\ref{graphic differential cross section}, the amount of backscattering is uniquely related to the chosen boundary condition. By measuring the amplitude of the backscattered wave, one is able in principle to determine which boundary condition has been specified by Nature.

\section{Stability}
\label{seciv}

All we did so far would be less relevant if the spacetime happens to be highly unstable. In view of that, our final task is to analyze how the stability of the system depends on the boundary condition. To do so, we recall that global monopoles allow for the existence of bound states only when $\beta>0$. For $\beta\le 0$ and $\beta=\pm\infty$, only scattering modes $\omega^2>0$ are allowed and, therefore, the system is mode stable. If $\beta>0$, on the other hand, the $\ell=0$ case admits a bound state of the form~(\ref{Rbound}), so that an arbitrary solution of the Klein-Gordon equation has to include not only the scattering states but also these bound states.

More precisely, the most general scalar wave can be decomposed as
\begin{align}
\Psi_\beta(t,r,\theta,\varphi)= A \frac{e^{-r/\beta}}{r}e^{-t/\beta}+B \frac{e^{-r/\beta}}{r}e^{t/\beta} \nonumber \\
+ \sum_{\ell=0}^{\infty}{\sum_{m=-\ell}^{\ell}{\int_{-\infty}^{\infty}{d\omega e^{-i\omega t} C_{\omega \ell m}{Y_\ell^m(\theta,\varphi)R_{\omega \ell}^{\beta}(r)}}}},
\label{complete solution}
\end{align}
where $A$, $B$ and $C_{\omega \ell m}$ are constants. The first term in the expansion above decays exponentially in time, becoming irrelevant after a sufficient long time (of order $\beta$). 
The second term of \eqref{complete solution}, however, grows exponentially in time. Nonetheless it still corresponds to a square-integrable solution because, for a fixed time $t$, the integral 
\begin{equation}
\int_{0}^{\infty}{\left|\frac{e^{-(r-t)/\beta}}{r}\right|^2r^2dr}=\frac{\beta}{2}e^{2t/\beta}
\end{equation}
is finite. Since this mode represents a growing perturbation, after a sufficient long time (of order $\beta$), non-linear effects will become important. While these effects may be able to control the exponential growth and restore the stability of the system, a full non-linear treatment of the Einstein-Klein-Gordon equations would be required to investigate that. For now, what we can say is that test scalar fields on the global monopole are mode unstable for $\beta>0$.

\section{Final Remarks}
\label{concluions}
We have seen that the propagation of scalar waves around a global monopole is not determined until a boundary condition at the origin is prescribed. This characterizes the global monopole as a quantum mechanically singular spacetime. The propagation of waves is as uncertain as the evolution of point particles reaching the classical singularity  at $r=0$. Assuming Nature has a preferred physical evolution scheme, this could be, as we discuss, inferred phenomenologically, allowing us to identify the boundary condition, for theory by itself is unable to predict it.

It is important to mention that the necessity of a boundary condition is due to the idealization of the global monopole's core. If we do not neglect its finite size, the boundary condition can, in principle, be related to the way the core is modelled (see, for instance, Ref.~\cite{kay}). Thus, perhaps a more physical and less mathematical way to interpret the main results of our analysis is that the differential cross section, instead of being related to the boundary condition chosen by Nature, is related to the specific details of the monopole's core.

In more detail, our analysis shows that the scattering amplitude and the differential cross section encode the arbitrariness of the boundary condition. In particular, the amount of backscattering is intimately related to the boundary condition parameter $\beta$. Consequently, in principle, by measuring the amplitude of a wave which is scattered by the global monopole (specially the backscattered part), one could determine the boundary condition in a given experiment (and, according to the reasoning above, extract information about the monopole core). 

Another important question we address in this paper concerns the stability of the global monopole. As we have shown, its stability under scalar perturbations depend on the boundary condition parameter. For $\beta>0$, the spacetime is unstable while for $\beta\le 0$ and $\beta=\pm\infty$, the spacetime is mode stable under perturbations of a test scalar field. Note, however, that the final word on the stability of the global monopole background requires a fully non-linear treatment of the problem. We also remark that the scalar test field under consideration here is not the same as the scalar fields $\phi^a$ that determine the global monopole background through \eqref{lagrangian}. The analysis developed in Ref.~\cite{achucarro} involving perturbations of such fields has shown no instabilities. 

Finally, the ideas here presented can, in theory, be extended to other naked singularity spacetimes, like the cosmic string, the negative-mass Schwarzschild, the overcharged Reissner-Nordstr\"om, and the overspinning Kerr spacetimes. The major technical difficulty in these cases, however, is to determine the self-adjoint extensions for the spatial part of the wave operator.  

\acknowledgments

We thank Alberto Saa for reading the manuscript and clarifying fundamental aspects of our work. The authors acknowledge support from the Sao Paulo Research Foundation (FAPESP) Grant No.~2013/09357-9. V.~S.~B.~and J.~P.~M.~P.~acknowledge support from FAPESP Grant No.~2016/08862-0. J.~P.~M.~P.~also acknowledges support from FAPESP Grant No.~2016/07057-6 and from FAEPEX Grant No.~2693/16. M.~R.~acknowledges support from the Fulbright Visiting Scholars Program, and is also grateful to Emanuele Berti and the University of Mississippi for hospitality while part of this research was being conducted.


\begin{thebibliography}{99}
\bibitem{vilenkin1}
A. Vilenkin, Phis. Rep. {\bf 121}, 263 (1985).

\bibitem{vilenkin2}
A. Vilenkin and E. P. S. Shellard, {\it Cosmic Strings and Other Topological Defects} (Cambridge University Press, Cambridge, England, 1994).

\bibitem{barriola}
M. Barriola and A. Vilenkin, Phys. Rev. Lett. {\bf 63}, 341 (1989).

\bibitem{bezerra de mello}
E.R. Bezerra de Mello, Braz. J. Phys. {\bf 31}, 211 (2001).

\bibitem{wald}
A. Ishibashi and  R. M. Wald, Class. Quant. Grav. {\bf 20}, 3815 (2003).



\bibitem{horowitz}
G. T. Horowitz and D. Marolf, Phys. Rev. D {\bf 52}, 5670 (1995).

\bibitem{mosna}
R. A. Mosna, J. P. M. Pitelli and M. Richartz, Phys. Rev. D {\bf 94}, 104065 (2016).

\bibitem{mazur}
P. O. Mazur and J. Papavassiliou, Phys. Rev. D {\bf 44}, 1317 (1991).

\bibitem{anacleto}
M.A. Anacleto, F.A. Brito, S. J. S. Ferreira and E. Passos, arXiv:1701.08147 [hep-th].


\bibitem{perivolaropoulos}
L. Perivolaropoulos, Nucl. Phys. {\bf B}, 665 (1992).


\bibitem{achucarro}
A. Ach\'ucarro and J. Urrestilla, Phys. Rev. Lett. {\bf 85}, 3091 (2000).

\bibitem{pitelli}
J. P. M. Pitelli and P. S. Letelier, Phys. Rev. D {\bf 80}, 104035 (2009).


\bibitem{ishibashi}
A. Ishibashi and A. Hosoya, Phys. Rev. D {\bf 60}, 104028 (1999).

\bibitem{deser}
S. Deser and R. Jackiw, Commun. Math. Phys. {\bf 118}, 495 (1988).

\bibitem{barroso}
V. S. Barroso and J. P. M. Pitelli, Phys. Rev. D{\bf 96}, 025006 (2017).

\bibitem{gitman}
D. M. Gitman, I. V. Tyutin and  B. L. Voronov, Jour. Phys. A {\bf 43}, 145205 (2010).

\bibitem{cotaescu}
I. I. Cotaescu, C. Crucean and  C. A. Sporea, Eur. Phys. J. C {\bf 76}, 102 (2016).

\bibitem{yennie}
D. R. Yennie, D. G. Ravenhall, and R. N. Wilson, Phys. Rev. {\bf 95}, 500 (1954).


\bibitem{kay}
B. Allen, B. S. Kay and A. C. Ottewill, Phys. Rev. D {\bf 53}, 6829 (1996).

\end{thebibliography}
\end{document}